\documentclass[pra,twocolumn,superscriptaddress,amsmath]{revtex4}
\usepackage{amssymb}
\usepackage{epsfig}
\begin{document}

\date{June 21, 2002}

\def\horparallel{
\begin{picture}(7,5)
\thicklines
\put(1,0){\line(1,0){5}}
\put(1,5){\line(1,0){5}}
\end{picture}}

\def\vertparallel{
\begin{picture}(8,6)
\thicklines
\put(1,0){\line(0,1){5}}
\put(6,0){\line(0,1){5}}
\end{picture}}

\def\Re{{\rm Re}}
\def\Im{{\rm Im}}
\def\bv{{\big\vert}}

\title{Ground-state properties of the Rokhsar--Kivelson  dimer model \\
on the triangular lattice}

\affiliation{Landau Institute for Theoretical Physics, 117940 Moscow,
Russia}
\affiliation{Theoretische Physik, ETH-H\"onggerberg, CH-8093
Z\"urich, Switzerland}

\author{A.~Ioselevich}
\affiliation{Landau Institute for Theoretical Physics, 117940 Moscow,
Russia}
\author{D.~A.~Ivanov}
\affiliation{Theoretische Physik, ETH-H\"onggerberg, CH-8093
Z\"urich, Switzerland}
\author{M.~V.~Feigelman}
\affiliation{Landau Institute for Theoretical Physics, 117940 Moscow,
Russia}

\begin{abstract}
We explicitly show that the Rokhsar--Kivelson dimer model on the
triangular lattice is a liquid with topological order. 
Using the Pfaffian technique, we prove that the difference
in local properties between the two topologically degenerate 
ground states on the cylinders and on the tori decreases 
exponentially with the system size. We compute the
relevant correlation length and show that it equals the
correlation length of the vison operator.
\end{abstract}

\maketitle


\def\Figureone{
\begin{figure}[b]
\epsfxsize=0.9\hsize
\centerline{\epsfbox{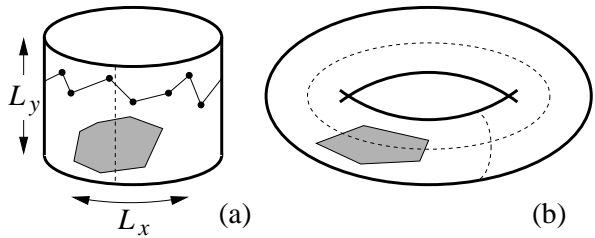}}
\medskip
\caption{
Topological sectors on a cylinder and on a torus. Dimer coverings
may be classified according to the parities of the number of dimers
intersecting the reference lines (dashed lines). Dimer configurations
differing by a rearrangement in contractible domains 
(shaded areas) belong to the same topological sector and contribute
with the same sign to the partition function (\ref{pfaffian}).
To change the topological sector, a circular permutation of dimers
along a topologically nontrivial contour (shown in a zig-zag line
in the case of the cylinder) is necessary.
}
\label{fig-cyltorus}
\end{figure}
}

\def\Figuretwo{
\begin{figure}[t]
\epsfxsize=0.7\hsize
\centerline{\epsfbox{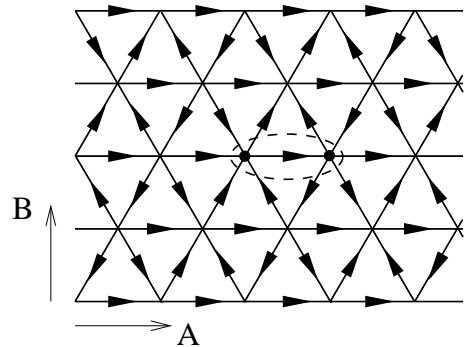}}
\medskip
\caption{
One possible choice of the amplitudes $A_{ij}$. This choice
of amplitudes is periodic with a unit cell containing two
lattice sites (marked by a dashed ellipse). The arrow
directions correspond to the signs of $A_{ij}$:
$A_{ij}$ equals 1 if the arrow points from $i$ to $j$, and
equals $-1$ if it points from $j$ to $i$.
Also shown are $A$ and $B$ directions: parallel and perpendicular 
to the lattice lines, respectively. 
}
\label{fig-lattice}
\end{figure}
}

\def\Figurethree{
\begin{figure}[bt]
\epsfxsize=0.8\hsize
\centerline{\epsfbox{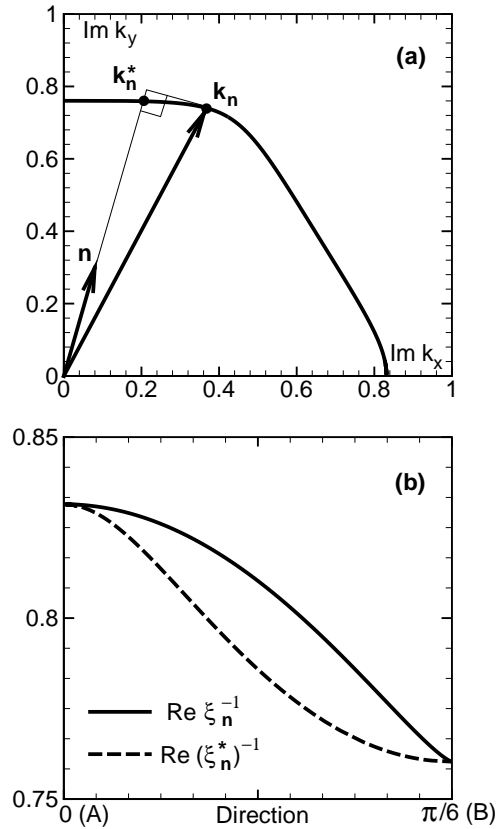}}
\caption{
Correlation length and its direction dependence.
{\bf (a)} Construction of the correlation length. 
In the $(\Im\, k_x, \Im\, k_y)$ plane, there is a region
around $\Im \, {\bf k}=0$ where there are no solutions to
the equation $E({\bf k})=0$. The thick solid line denotes
the boundary of this region [the axis $k_x$ is chosen along
one of the lattice directions ($A$ direction), and the axis $k_y$ is 
perpendicular to it ($B$ direction);
if reflected in those two axes, the thick solid line has the
hexagonal symmetry]. Given the direction ${\bf n}$, the
vector $\Im\,{\bf k}_{\bf n}$ 
is chosen on this boundary to maximize its projection onto ${\bf n}$
[in accordance with (\ref{condition-1}),
(\ref{condition-2}), or, equivalently, with (\ref{condition-min})]
The vector ${\bf k}_{\bf n}$ determines the correlation length
$\xi_{\bf n}^{-1} = -i {\bf n} {\bf k}_{\bf n}$ 
in the ${\bf n}$ direction. For a vison tunneling as
a wave front [Eq.~(\ref{torus-splitting-three})], 
the tunneling amplitude has a different correlation length given by 
$(\xi^\star_{\bf n})^{-1} = -i {\bf n} {\bf k}_{\bf n}^\star$ 
with the imaginary component of the vector
${\bf k}_{\bf n}^\star$
parallel to ${\bf n}$.
{\bf (b)} The direction dependence of the real parts of the
inverse correlation lengths
$\xi_{\bf n}^{-1}$ and $(\xi^\star_{\bf n})^{-1}$. The inverse
correlation lengths are plotted versus the angle between the direction
${\bf n}$ and the lattice lines. The angle $0$ corresponds to the
$A$ direction, and the angle $\pi/6$ corresponds to the $B$ direction. 
}
\label{fig-xi}
\end{figure}
}

\def\Figurefour{
\begin{figure}[tb]
\epsfxsize=0.6\hsize
\centerline{\epsfbox{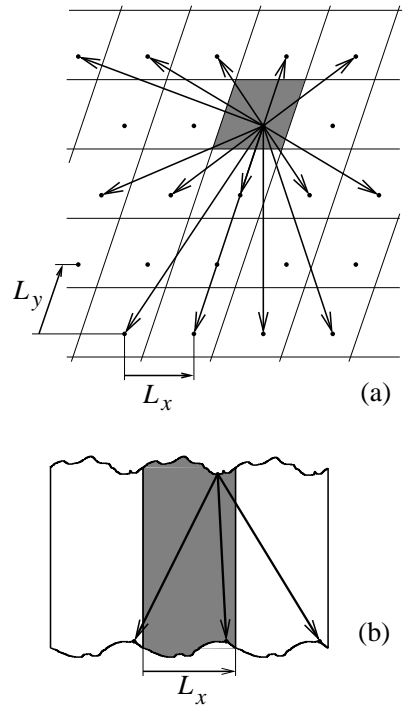}}
\medskip
\caption{The vison tunneling trajectories ${\bf R}$ contributing to
the splitting between partition functions on the torus and on the
cylinder.
{\bf (a)} 
The torus is constructed by identifying points on the plane
differing by multiples of the two basis vectors ${\bf L}_x$
and ${\bf L}_y$. Equivalently, it may be described by identifying
opposite sides of the unit cell (shaded parallelogram).
The vectors ${\bf R}$ contributing to
the splitting between $Z_{++}$ and $Z_{+-}$
(or, equivalently, between $N_{eo}$ and $N_{oo}$)
[Eqs.~(\ref{Z-splitting}) and (\ref{torus-splitting-one})]
belong to the lattice generated by ${\bf L}_x$
and ${\bf L}_y$ and have odd winding numbers in the $y$ direction.
{\bf (b)} The tunneling trajectories ${\bf R}$
determining the splitting between the topological sectors 
on the cylinder as given by Eq.~(\ref{torus-splitting-one}).
The cylinder is defined by identifying points differing by
a multiple of ${\bf L}_x$. 
}
\label{fig-split}
\end{figure}
}

\def\Figurefive{
\begin{figure}[bt]
\epsfxsize=1.0\hsize
\centerline{\epsfbox{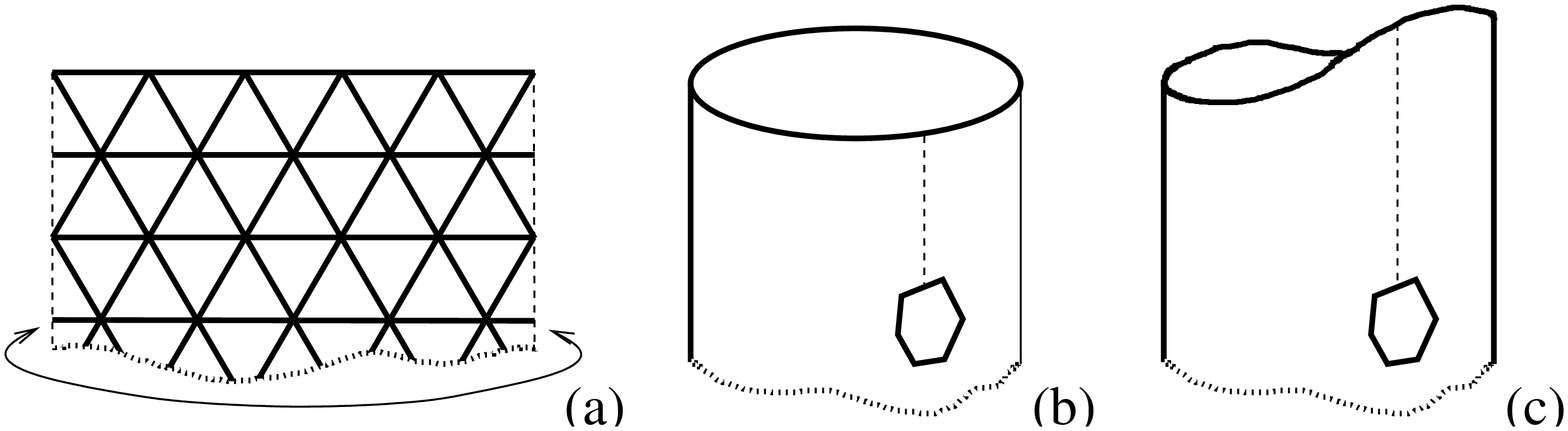}}
\medskip
\caption{
{\bf (a)} A particular geometry of the semi-infinite cylinder
with the straight edge in the $A$ direction. For this cylinder,
the zero mode of the $A_{ij}$ matrix (for one of the two boundary
conditions) may be explicitly found. Arrows show identification
of the two edges of the semi-infinite stripe.
{\bf (b)} The same cylinder as in (a), with a hole sufficiently
far from the edge. Two different boundary conditions are possible across the
dashed line.
{\bf (c)} A cylinder of arbitrary geometry with a hole far from the
edge. Two different boundary conditions are possible across the
dashed line.
}
\label{fig-cylinders}
\end{figure}
}

\def\Figuresix{
\begin{figure}[b]
\epsfxsize=0.6\hsize
\centerline{\epsfbox{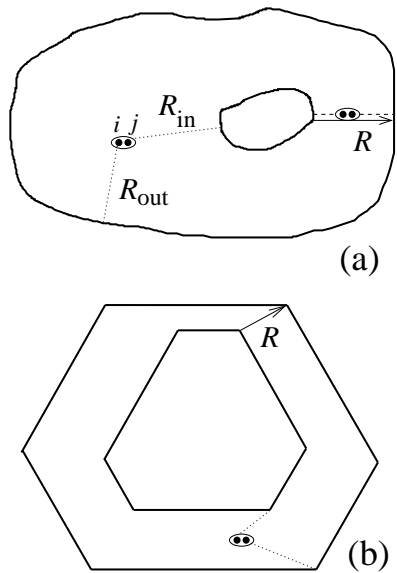}}
\medskip
\caption{
The vison tunneling trajectories determining the splitting between the
topological sectors in the plane geometry (a disc with a hole).
{\bf (a)} In the general situation, the splitting is determined by
Eq.~(\ref{torus-splitting-one}) with the minimization over trajectories
${\bf R}$ connecting the hole
to the external boundary. The splitting of the local dimer correlation
functions (\ref{correlation-splitting-two}) 
is determined by the two tunneling events: vison
tunneling from the inner boundary to the location of the involved
dimers ($R_{\rm in}$) and then further 
to the outer boundary ($R_{\rm out}$).
The two possible situations are shown: the splitting at locations near the
optimal tunneling trajectory have the same exponential dependence
as the direct tunneling from the inner to the outer boundary (dashed line),
while at locations far from the optimal tunneling trajectory, the
splitting is smaller (dotted line).
{\bf (b)} In the special case of polygon boundaries with their 
segments parallel to
lattice directions ($A$ directions), only trajectories connecting
corners of the boundaries should be taken into account in
(\ref{torus-splitting-one}) and (\ref{correlation-splitting-two}),
since vison tunneling from the boundaries is suppressed.
}
\label{fig-plane}
\end{figure}
}

\def\Figureseven{
\begin{figure}[bt]
\epsfxsize=0.8\hsize
\centerline{\epsfbox{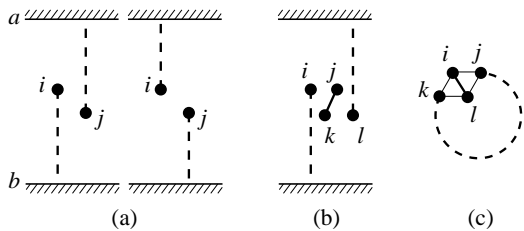}}
\medskip
\caption{
Diagrams involved in splitting between correlation functions
in different topological sectors.
{\bf (a)}
The two diagrams contributing to the splitting of the average
dimer density $\langle n_{ij}\rangle$ between the odd and even 
sectors on the cylinder. The two edges of the cylinder are labeled
$a$ and $b$. The dashed lines represent the zero-mode wave functions
$\Psi_a$ and $\Psi_b$.
{\bf (b)}
One of the diagrams contributing to the splitting of the four-point
correlation function $\langle n_{ij} n_{kl} \rangle$ on the cylinder.
The dashed lines represent  the zero-mode wave functions
$\Psi_a$ and $\Psi_b$; the solid line denotes the bulk Green's function
$G_{jk}$.
{\bf (c)}
One of the diagrams contributing to the splitting of the four-point
correlation function $\langle n_{ij} n_{kl} \rangle$ on the torus.
The thick solid line denotes the bulk  Green's function
$G_{il}$. The dashed line denotes the Green's function with a winding
around the torus $G_{j,k+{\bf R}}$, where the winding vector
${\bf R}$ is one of those shown in Fig.~\ref{fig-split}.
}
\label{fig-diagrams}
\end{figure}
}

\def\Figureeight{
\begin{figure}[b]
\epsfxsize=0.5\hsize
\centerline{\epsfbox{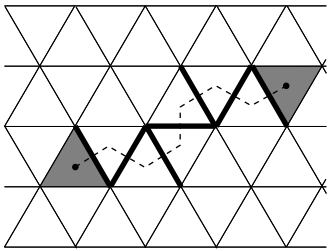}}
\medskip
\caption{
The product of the two vison operators. The visons are placed
in the shaded triangles. The contour $\Gamma$ connecting them
is shown as the dashed line. The bold links are those involved
in the operator (\ref{vison-vison}).
}
\label{fig-visons}
\end{figure}
}

\def\Figurenine{
\begin{figure}[bt]
\epsfxsize=0.9\hsize
\centerline{\epsfbox{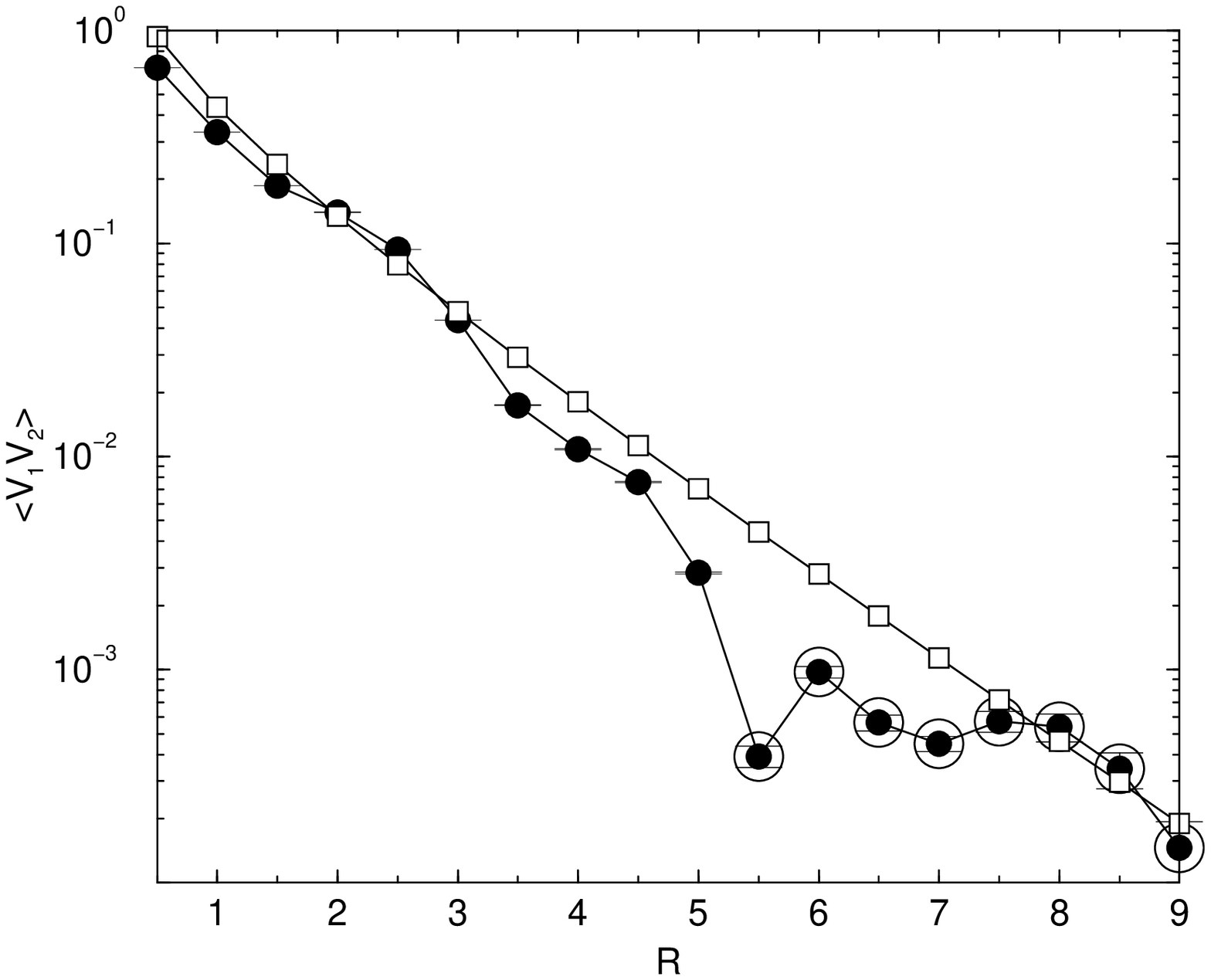}}
\medskip
\caption{
The variational Monte Carlo results of the vison-vison
correlation function (\ref{vison-vison})
in the $A$ direction. The plot is
in the linear-logarithmic scale, with the distance
measured in the lattice spacings. The solid dots
are the numerical results (the circles around dots indicate
negative values). The magnitude of the numerical data
should be compared to the analytic expression
$R^{-1/2}\exp(-\Re\, R/\xi_A)$ (empty squares), where
$\xi_A$ is given by (\ref{xi-A}). The calculation was
performed on the 50$\times$50 torus with averaging over
$10^6$ dimer configurations (error bars are shown).
}
\label{fig-numerics}
\end{figure}
}


\section{RVB liquid and the Rokhsar--Kivelson dimer model}

Resonating valence bond (RVB) spin liquid in two dimensions
is a remarkable theoretical concept which predicts very
unusual low-temperature properties of spin-1/2 systems\cite{PWA}. Unlike
conventional magnetically-ordered ground states with spin-1
excitations, the RVB ground state has no long-range order of any
local order parameter and possesses elementary excitations with
spin 1/2\cite{Kivelson-Rokhsar-Sethna}. 
If such a system is doped with mobile holes, the
fractionalization of spin excitations translates into the effect
of spin-charge separation: this scenario has been widely explored
in the context of high-temperature 
superconductivity\cite{Kivelson-Rokhsar-Sethna,Fradkin-Kivelson,%
Senthil-Fisher}. 
While a rigorous verification
of the RVB liquid phase in a realistic spin system is usually very 
difficult due to the strongly-correlated nature of the state,
many specially designed systems have confirmed RVB liquid
properties\cite{Lhuillier,Moessner-Sondhi,Balents-Fisher-Girvin,%
Senthil-Motrunich,Misguich-Serban-Pasquier,Ioffe-Feigelman}.

In understanding generic properties of the RVB liquid state, one
may benefit from studying dimer models which are closely related
to the RVB spin liquids\cite{Moessner-Sondhi-Fradkin,Read-Sachdev}.  
In the RVB construction, the wave function
is represented as a sum over singlet configurations, while in the
dimer models the singlets are replaced by dimers. The difference between
the spin and the dimer systems is that dimer configurations
are mutually orthogonal by definition, while different singlet configurations
have a finite overlap\cite{LDA}. The RVB spin liquid is known to have two 
types of elementary excitations: spinons (spin-1/2 excitations) and
visons ($Z_2$ vortices)~\cite{Senthil-Fisher,Read-Chakraborty}. 
In a dimer liquid,
the spinons are prohibited by the dimer constraint (or, equivalently,
are pushed infinitely high in energy), while the visons are expected to be
indeed the lowest excitations above the ground state. The RVB spin liquid
must have topological degeneracy on domains of nontrivial 
topology.
Such a decoupling into topological sectors is straightforward 
in dimer liquids where it is defined in purely geometric 
terms\cite{Read-Chakraborty,Rokhsar-Kivelson,Bonesteel}. 
Thus dimer models provide a convenient test ground for studying
properties of RVB liquids related to vison excitations.

The advantage of studying dimer models is that they are simpler than
spin models, and are hence better understood and more accessible to
analytical methods. The simplest dimer model contains the pair-wise hopping
term and the pair-wise potential term
\begin{equation}
H=\sum_\square \left[-t \left(
\left| \horparallel  \right\rangle\left\langle \vertparallel \right|+
\left| \vertparallel \right\rangle\left\langle \horparallel  \right| \right)
+v \left(
\left| \horparallel  \right\rangle\left\langle \horparallel  \right|+
\left| \vertparallel \right\rangle\left\langle \vertparallel \right| \right)
\right]\, ,
\label{dimer-hamiltonian}
\end{equation}
where the sum is taken over the four-vertex plaquets of the 
lattice\cite{Rokhsar-Kivelson}.
On the square lattice such a model has a crystal ground state (with
the crystal of dimers breaking the translational symmetry of the
lattice) for any ratio $v/t$, except for the phase-transition point(s) between
different crystal 
structures\cite{Moessner-Sondhi-Fradkin,Rokhsar-Kivelson,Levitov}. 
The situation is different on the
triangular lattice (with the sum in (\ref{dimer-hamiltonian}) taken
over all rhombi consisting of two neighboring triangles):
there the dimers are believed to form a liquid in a finite range
of the parameter $v/t$\cite{Moessner-Sondhi,Ioffe}. 
Out of all possible values of $v/t$, one
is special: $v/t=1$. At this ratio of parameters --- further called
Rokhsar--Kivelson (RK) point --- the ground state of the Hamiltonian
(\ref{dimer-hamiltonian}) is known exactly: it is a superposition of
all possible dimer configurations with equal amplitudes [unfortunately,
the excited states are not known exactly even at the RK point].
At $v/t>1$, this state immediately yields to the staggered crystal 
state via a first-order phase transition. However at $v/t<1$, 
from the available numerical evidence it follows that the RK state 
on the triangular lattice smoothly evolves without crystallization
in a finite range of $v/t$ (approximately until $v/t \approx 0.6 \dots 0.8$)%
\cite{Moessner-Sondhi,Ioffe}.
It is this region of the parameter space which contains the dimer liquid.

The RK point is one of the representatives of the liquid phase and
is most accessible for analytic treatment, because averaging over the
ground state is equivalent to the statistical averaging over all
possible dimer configurations. Such an averaging may be performed
with the usual Pfaffian technique\cite{Kasteleyn}
which allows to establish the exponential decay of dimer
correlations\cite{Moessner-Sondhi}.
In this paper, we employ the
Pfaffian technique to explicitly compute the correlation length
involved in the ground-state correlation functions and in the
splitting between the topological sectors on the cylinder and on the
torus. We further demonstrate that
the same correlation length also appears in the vison correlation
function. Thus our calculations prove that the RK ground state 
on the triangular lattice possesses topological order: on a topologically
nontrivial domain, the ground states in different topological sectors
are degenerate
and locally indistinguishable (in the limit of infinite system size). 
Note that the exponential decay of ground-state correlation 
functions is consistent with the numerical finding of the gap in
the excitation spectrum 
at the RK point (the magnitude of the gap is numerically estimated as 
$0.1 t$)\cite{in-preparation,Ioffe}.

\section{Pfaffian technique and correlation lengths}
\label{section-pfaffian}

In the Pfaffian technique, the statistical averaging over the dimer 
configurations is performed via introducing an auxiliary real fermionic
variable (a Majorana fermion) on each lattice site\cite{Kasteleyn}. 
The total number
of dimer configurations may then be written as  the partition function
of these fermions
\begin{equation}
Z=\int \prod_i da_i \, \exp\bigg[\sum_{ij} a_i A_{ij} a_j\bigg]
={\rm Pfaff}(A_{ij}),
\label{pfaffian}
\end{equation}
where $i$ and $j$ label the lattice sites, and the fermionic variables
$a_i$ obey the conventional rules: $a_i a_j=-a_j a_i$, $\int da_i=0$,
$\int a_i\, da_i=1$. The ``hopping amplitudes'' $A_{ij}$ take values
$\pm 1$ on nearest-neighbor sites and $0$ otherwise, and form an 
antisymmetric matrix: $A_{ij}=-A_{ji}$. The signs of $A_{ij}$ must be
adjusted so that all terms in the expansion of the exponent in
(\ref{pfaffian}) give
positive weight. Such terms in the expansion of the exponent are in
one-to-one correspondence to the dimer coverings of the lattice.
Each of them has magnitude one, and should they all have equal signs,
the partition function (\ref{pfaffian}) simply counts the total number of
dimer coverings.

\Figureone

As shown in Ref.~\onlinecite{Kasteleyn}, a necessary requirement for the
proper relative sign of different dimer coverings is that the circular
product $\prod A_{ij}$ equals $(-1)$ around any elementary rhombus
(around any elementary even-length cycle, 
in the general case of a planar lattice).
Then it follows that $\prod A_{ij}=-1$ over any {\it contractible}
contour of even length and, as a consequence, any two dimer configurations
differing by a {\it local} rearrangement of dimers contribute to the
partition function (\ref{pfaffian}) with equal signs. Different dimer
configurations which cannot be related by local dimer rearrangements
(i.e.\ by rearrangements in a contractible domain, 
see Fig.~\ref{fig-cyltorus})
may contribute with either equal or opposite signs: this effect will
be of crucial importance for our calculation.

\Figuretwo

One possible choice of the amplitudes $A_{ij}$ is shown in 
Fig.~\ref{fig-lattice}.
Note that in order to obey the sign rule for $A_{ij}$
we have to double the unit cell of the lattice. However, any physical
quantity computed with those $A_{ij}$ has the periodicity of the lattice,
as the set of amplitudes $A_{ij}$ translated by one lattice spacing 
may be returned to the original form by an appropriate $Z_2$ gauge
transformation $A_{ij}\mapsto W_i A_{ij} W_j$ with $W_i=\pm 1$.
Now the partition function (\ref{pfaffian}) as well as correlation
functions may be conveniently computed in the Fourier components.
The Fourier transformation of $A_{ij}$ is a $2{\times}2$ matrix
$A({\bf k})$. 
From antisymmetricity of the matrix $A_{ij}$ it follows that its
eigenvalues always come in pairs $\pm i E({\bf k})$, with
complex conjugate eigenvectors corresponding to opposite eigenvalues.
A straightforward calculation shows that, for the triangular lattice, 
the spectrum of $A({\bf k})$ in the bulk is always gapped 
(never crosses zero as a function of ${\bf k}$). Explicitly, 
it is given by
\begin{equation}
E({\bf k})=2 \left( \cos^2 k_1 + \cos^2 k_2 + \cos^2 k_3 \right)^{1/2},
\label{spectrum}
\end{equation}
where $k_1$, $k_2$, and $k_3$ are the projections of the vector ${\bf k}$
onto the three lattice directions (obeying the
constraint $k_1+k_2+k_3=0$). We have also shifted the origin of the
Brillouin zone to bring (\ref{spectrum}) to a symmetric form.
The gap in the spectrum of $A_{ij}$ is crucial for the exponential decay of
correlation functions: as we shall see below, the correlation length
is determined by the complex wave vectors ${\bf k}$ solving the
equation $E({\bf k})=0$\cite{Moessner-Sondhi}. 
At this stage of calculation, we see the
difference between the triangular and square lattices: on the square
lattice, the corresponding matrix $A_{ij}$ leads to a gapless spectrum
(and, consequently, to the power-law 
decay of correlations\cite{Fisher-Stephenson}.
Note that although the presence or absence of the gap in the matrix
$A_{ij}$ (governing the decay of correlation functions in the ground
state) agrees with
the presence or absence of the gap for excitations, there is no direct 
relation between the two gaps. The spectrum of $A_{ij}$ is not the
energy spectrum of quantum dimers, but only the spectrum of auxiliary
Majorana fermions introduced for calculating the classical partition
function of all dimer configurations.

Correlation functions of dimers may be expressed in terms of Green's
functions of Majorana fermions. Indeed, the probability of dimers
occupying a given set of positions (connecting pairwise points
1 and 2, 3 and 4, \dots, $2m-1$ and $2m$) may be computed by excluding
those points from the lattice, which is equivalent to placing
Majorana fermions at those points:
\begin{multline}
\langle n_{12} \dots n_{2m-1,2m}\rangle =
\langle a_1 a_2 \dots a_{2m} \rangle \\
\equiv
Z^{-1} \int \prod_i da_i \, (a_1 a_2 \dots a_{2m})\, 
\exp\bigg[\sum_{ij} a_i A_{ij} a_j\bigg]
\label{multi-point}
\end{multline}
(up to a sign).
Here the first average denotes the statistical averaging of
dimer occupation numbers $n_{ij}$ over
all dimer configurations, while the second average is the
quantum-mechanical correlation function in the theory of Majorana
fermions. This correlation function may further be decoupled using
the Wick theorem.

The Green's functions are obtained as
\begin{equation}
G(i,j)\equiv\langle a_i a_j \rangle =
\int_{\rm B.Z.} d{\bf k}\, A^{-1}({\bf k}) e^{i{\bf k}({\bf r}_i-{\bf r}_j)},
\label{Majorana-Green}
\end{equation}
with the integral over ${\bf k}$ defined as averaging over the
Brillouin zone. The decay at large distances is found from deforming
the integration domain into complex values until it reaches the
singularities in $A^{-1}({\bf k})$. The singularities are the square-root
branching points, which translates into the asymptotics
$G(R) \propto R^{-1/2} \exp(-R/\xi)$. The correlation length $\xi$
is direction-dependent. In the direction specified by a unit vector ${\bf n}$,
the correlation length $\xi_{\bf n}$ is given by 
$\xi_{\bf n}^{-1}= -i {\bf k}_{\bf n} {\bf n}$,
where the complex vector ${\bf k}_{\bf n}$ is determined
from the set of equations:
\begin{align}
E({\bf k}_{\bf n}) &= 0; 
\label{condition-1}\\
\nabla E({\bf k}_{\bf n}) &\parallel {\bf n}. 
\label{condition-2}
\end{align}
(the meaning of the second equation is that the ``group velocity'' is directed
along the vector ${\bf n}$). Equivalently, the 
conditions (\ref{condition-1}), (\ref{condition-2}) for ${\bf k}_{\bf n}$
may be expressed as
\begin{equation}
\xi_{\bf n}^{-1} =\max\min\nolimits^\Re \,
(-i {\bf k} {\bf n}) \, \bv_{E({\bf k})=0},
\label{condition-min}
\end{equation}
where $\max\min\nolimits^\Re$ denotes minimizing the
real part (among its positive values) over the components
of ${\bf k}$ parallel to ${\bf n}$ and maximizing it
over the perpendicular components.
The vectors ${\bf k}$ are confined to
the complex surface $E({\bf k})=0$.
The construction of $\xi_{\bf n}^{-1}$ is illustrated in
Fig.~\ref{fig-xi}.

\Figurethree

In the particular cases of 
directions $A$ and $B$ parallel and perpendicular to the lattice lines
respectively (as shown in Fig.~\ref{fig-lattice}),
the correlation lengths may be found analytically 
(from solving (\ref{condition-1}) and (\ref{condition-2}))
\begin{align}
\xi_A^{-1} &= {1\over2} \ln \left(1+\sqrt3+\sqrt{2\sqrt3+3}\right) 
\pm i \arctan\sqrt{{2\over\sqrt3}-1} \nonumber\\
& \approx 0.83 \pm 0.12 \, i\pi \, , 
\label{xi-A}\\
\xi_B^{-1} &= {1\over\sqrt3} \ln \left(2+\sqrt3\right)\approx 0.76 
\label{xi-B}
\end{align}
(in the units of the lattice constant). 
The imaginary part in $\xi_A^{-1}$
implies the damped oscillating asymptotic behavior (with an
incommensurate wave vector) of the Green's function and
of other relevant correlations (e.g., vison correlation function,
see Section~\ref{section-vison}).

For other directions, the equations (\ref{condition-1}) and 
(\ref{condition-2}) [or, equivalently, (\ref{condition-min})]
may be solved numerically. In Fig.~\ref{fig-xi}, we plot
the direction dependence of the inverse coherence length
$\xi_{\bf n}^{-1}$.

The exponential decay of
the Green's functions implies an exponential decay of all dimer-dimer
correlations. For example, the pair-wise dimer correlation
$\langle n_{12} n_{34}\rangle$ decays with the correlation length $\xi/2$,
since it involves a product of two Green's functions\cite{Moessner-Sondhi}.

\section{Topological degeneracy on multiply-connected domains}
\label{section-degeneracy}

If one considers dimer coverings on a domain of nontrivial topology
(e.g., torus, cylinder, domains with holes, etc.), such coverings
may be classified into topological classes\cite{Read-Chakraborty,Bonesteel}. 
Dimer coverings
from different topological classes cannot be transformed into each
other by local rearrangements of dimers. The topological classes are
defined by the parity of the number of dimers intersecting a given
non-contractible closed contour. On the cylinder there are two topological
classes, on the torus there are four of them (Fig.~\ref{fig-cyltorus}).

In the quantum dimer model, the Hilbert spaces spanned by different 
topological classes are not mixed by the Hamiltonian, and each topological
class possesses its own ground state. At the RK point, each of these
ground states has zero energy. In the general case of a topological liquid,
the splitting of energies between different topological sectors must
be exponentially suppressed in system size\cite{Senthil-Fisher,%
Moessner-Sondhi-Fradkin}. We verify this property
later near the RK point to the lowest-order perturbation theory in
$(v/t-1)$.

At the RK point, we compare correlation functions in different topological
sectors. We shall see that the correlators in different sectors nearly
coincide, with the splitting exponentially small in system size. We start
our analysis with comparing partition functions for different topological
sectors. The derivation looks different for the cases of the torus and 
of the cylinder.

Consider first the topological sectors on the torus. Denote the
numbers of dimer coverings with even/odd intersection indices at
reference contours by $N_{ee}$, $N_{eo}$, $N_{oe}$, and $N_{oo}$,
where $N_{\epsilon_x \epsilon_y}$ corresponds to the sector with
the parity $\epsilon_x$ of intersection with the line parallel to
$y$-axis, and the parity $\epsilon_y$ is of intersection with the
line parallel to $x$-axis.
In the Pfaffian technique, different topological sectors may be
accessed by imposing either periodic or antiperiodic boundary
conditions across the reference contours (in the $x$ and $y$ directions)%
\cite{Read-Chakraborty,Bonesteel,I-S}. 
The corresponding
partition functions $Z_{\sigma_x \sigma_y}$ ($\sigma_x,\sigma_y=\pm$)
are expressed in terms of $N_{\epsilon_x \epsilon_y}$ as
\begin{align}
Z_{++}&=N_{ee}+N_{eo}+N_{oe}-N_{oo}\, ; \nonumber\\
Z_{+-}&=N_{ee}-N_{eo}+N_{oe}+N_{oo}\, ; \nonumber\\
Z_{-+}&=N_{ee}+N_{eo}-N_{oe}+N_{oo}\, ; \\
Z_{--}&=-N_{ee}+N_{eo}+N_{oe}+N_{oo}\, . \nonumber
\end{align}
(up to the simultaneous change of signs of $\sigma_x$ and/or
$\sigma_y$ in all the four expressions above: the particular
choice of signs of $\sigma_x$ and $\sigma_y$ depends
on the gauge choice for $A_{ij}$;
the overall sign of $Z_{\sigma_x\sigma_y}$ is ignored,
as usual).
These expressions are a particular case of a more general
formula for surfaces of arbitrary genus\cite{Zecchina}.
We now establish that the splitting in $Z_{\sigma_x\sigma_y}$ 
is exponentially small in the system size, then the same
exponential smallness will follow for the splitting
of $N_{\epsilon_x \epsilon_y}$.
Consider, for example, the partition functions $Z_{++}$ and $Z_{+-}$.
Each of them is given by the sum over the Brillouin zone
\begin{equation}
Z_{\sigma_1\sigma_2}=\exp\, \sum_{\rm B.Z.} \ln E({\bf k}),
\end{equation}
where the lattice of points ${\bf k}$ is determined by the boundary conditions.
In the partition functions $Z_{++}$ and $Z_{+-}$, the lattices of ${\bf k}$ are
shifted by half a period in the $y$-direction. We can rewrite the
relative difference $\Delta$ 
in those partition functions as a sum over trajectories
with odd windings in the $y$-direction (taking the Fourier transform
of $\ln E({\bf k})$ and using the Poisson summation formula):
\begin{align}
\Delta &={Z_{++} - Z_{+-}\over Z}=
\sum_{\rm B.Z.} \ln E({\bf k}) - {\sum_{\rm B.Z.}}' \ln E({\bf k'})
\nonumber\\
&= \sum_{\bf R} f({\bf R}),
\label{Z-splitting}
\end{align}
where $f({\bf R})$ is the Fourier transform of $\ln E({\bf k})$, and the
sum is taken over all closed trajectories with odd windings
in the $y$-direction (see Fig.~\ref{fig-split}a). The function $f({\bf R})$
decays exponentially with distance, with the same correlation
lengths $\xi_{\bf n}$ as the Majorana Green's functions (the correlation
lengths determined by (\ref{condition-1}) and (\ref{condition-2})).
The physical interpretation of those exponents
are the quasiclassical amplitudes of the vison tunneling around
the torus\cite{Senthil-Fisher}.
The splitting (\ref{Z-splitting}) is dominated by the largest
of the exponents, i.e., by the optimal trajectory (the choice of the
optimal trajectory depends on the aspect ratio of the torus,
and, because of the anisotropy, may not necessarily be the 
geometrically shortest one).
This result may formally be written as
\begin{equation}
\Delta \lesssim 
\exp\left(-\min_{\bf R} \Re\, \left[R\over\xi_{\bf n}\right]\right),
\label{torus-splitting-one}
\end{equation}
where $\xi_{\bf n}$ is given by (\ref{condition-1}) and (\ref{condition-2})
[or, equivalently by (\ref{condition-min})]
with the vector ${\bf n}$ in the direction of ${\bf R}$, and the minimization
is performed over all displacements ${\bf R}$ returning to the
original point with an odd winding number in the $y$-direction
(Fig.~\ref{fig-split}a).

Note that Eq.~(\ref{torus-splitting-one}) neglects interference
between different trajectories in Fig.~\ref{fig-split}a. Such
an interference may additionally suppress the splitting, hence
we put the $\lesssim$ sign in (\ref{torus-splitting-one}).
We can take the interference into account by performing not the full
Fourier transformation in (\ref{Z-splitting}), but only the Fourier
transformation in the $y$ direction, leaving the $k_x$ component
of the wave vector quantized. We denote this quantization of
$k_x$ as $k_x\in\Lambda$, where $\Lambda$ is the set of allowed
values of $k_x$ [necessarily real!] depending on $L_x$ and on the
boundary condition. Then repeating the above argument, we arrive
at the estimate for the splitting:
\begin{equation}
\Delta \lesssim
\exp\left( - \min \Im \, {\bf k} {\bf L}_y\, 
\bv^{k_x\in\Lambda}_{E({\bf k})=0} \right),
\label{torus-splitting-two}
\end{equation}
where the minimum is taken among positive values only.
Because of the constraint $k_x\in\Lambda$, the expression
(\ref{torus-splitting-two}) is invariant with respect to
${\bf L}_y \mapsto {\bf L}_y \pm {\bf L}_x$, as expected
[any displacement ${\bf R}$ in Fig.~\ref{fig-split}a with winding
number one may be chosen as ${\bf L}_y$].

We have obtained the two estimates (\ref{torus-splitting-one}) and
(\ref{torus-splitting-two}) both derived in the limit $L_y\gg 1$.
One can verify that the estimate (\ref{torus-splitting-one})
is stronger than (\ref{torus-splitting-two}) when
$L_x \gtrsim L_y \gg 1$, while (\ref{torus-splitting-two})
is stronger than (\ref{torus-splitting-one}) when 
$L_y \gg L_x \sim 1$ (quasi-one-dimensional limit).
In the intermediate regime $L_y \gg L_x \gg 1$, both
(\ref{torus-splitting-one}) and (\ref{torus-splitting-two})
reduce to the same result
\begin{align}
\label{torus-splitting-three}
& \Delta \sim
\exp\left(- \Re\, \left[ L_y \over \xi^\star_{\bf n}\right] \right),
\\
& 
(\xi^\star_{\bf n})^{-1}= \min\nolimits^\Re\, (- i{\bf n}{\bf k})\,
\bv^{E({\bf k})=0}_{\Im {\bf L}_x {\bf k} =0 }. \nonumber
\end{align}
Here ${\bf n}$ is the vector perpendicular to ${\bf L}_x$ and
$L_y = {\bf n} {\bf L}_y$.
The minimization is performed over the wave vector ${\bf k}$ with
its imaginary component parallel to ${\bf n}$.
Thus defined correlation length 
$\xi^\star_{\bf n}$
corresponds to the vison tunneling as a ``wave front''
(as a plane wave in the $x$ direction) and is slightly
different from $\xi_{\bf n}$ corresponding to the
tunneling of a ``point'' vison, see Fig.~\ref{fig-xi}
(this difference is due to the lattice anisotropy).

The splitting between $Z_{++}$ and $Z_{+-}$ translates into the exponentially
small splitting between $N_{eo}$ and $N_{oo}$ given by the
same expressions 
(\ref{torus-splitting-one})--(\ref{torus-splitting-three}). 
Note that a vison
tunneling in the $y$-direction relates the topological sectors with 
different parity of intersections with a contour running in the 
$y$-direction. Similar expressions apply for splitting
in other pairs of topological sectors.

The splitting of the two topological sectors on a cylinder 
(or on a disc with one hole, which is topologically equivalent
to a cylinder) is derived
in a way slightly different from our analysis of the torus. 
Consider a semi-infinite cylinder (or a plane with one hole of
arbitrary size and shape). On such a system, we consider
two different boundary conditions for Majorana fermions:
periodic and antiperiodic as we go around the cylinder (around
the hole, respectively). To those boundary conditions, there correspond
the two matrices $A_{ij}^{(+)}$ and $A_{ij}^{(-)}$.
Below we prove that one of those matrices has a zero
eigenvalue (with the eigenvector localized near the boundary), and the
other does not (in the general case).

\Figurefour

The proof consists of four steps.
First, we consider a semi-infinite
cylinder of a particular geometry, with the straight boundary
parallel to one of the lattice lines (Fig.~\ref{fig-cylinders}a). 
For such a cylinder, the spectrum can be exactly
computed, and the existence of the zero mode in one of the two topological
sectors is easily verified (for this particular geometry, the zero
mode is strictly localized at the boundary row of sites). Second, 
we note that the bulk spectrum of $A_{ij}$ is gapped, 
and therefore there are only a finite number
of states at small energies, and they are all localized near the
boundaries. Since all eigenvalues of the antisymmetric
matrix come in pairs $\pm iE$, the property of having a zero mode depends
on whether there are even or odd number of subgap states.
This parity can not be changed by any {\it local} deformation
of the antisymmetric matrix $A_{ij}$ and thus the zero mode
(or its absence) is topologically stable (note that this
argument resembles the proof of the topological stability 
of the zero mode in vortices
in $p$-wave superconductors\cite{Read-Green,Iv}).
Third, once the theorem proven for a particular geometry of the
cylinder, we make a hole in this cylinder sufficiently far from
the boundary (Fig.~\ref{fig-cylinders}b). Then we can impose either
periodic or antiperiodic boundary conditions across a line connecting
the hole to the edge of the cylinder (dashed line in the figure).
Switching the boundary conditions toggles on and off the zero mode
at the cylinder edge. However, since the total parity of the
subgap states (including both states at the cylinder edge and
at the hole) is conserved, switching the boundary conditions 
across the dashed line also toggles the zero mode at the hole
boundary. Thus we extend our theorem to the case of a hole
in a plane (a sufficiently large cylinder may be regarded as a plane,
from the point of view of states localized near the hole).
Fourth, we repeat the previous argument for a cylinder of arbitrary
geometry with a hole in it (Fig.~\ref{fig-cylinders}c), and thus
establish the presence of the zero mode at the cylinder edge
for one of the two boundary conditions. This completes the proof.

\Figurefive

\Figuresix

It is the zero modes which determine the splitting between the topological
sectors on a cylinder. The partition functions of the Majorana fermions
on the cylinder with periodic and antiperiodic boundary conditions are
\begin{equation}
Z_+=N_e+N_o\, , \qquad Z_-=N_e-N_o
\end{equation}
(or vise versa, depending on the choice of the gauge for $A_{ij}$),
where $N_e$ and $N_o$ are the numbers of dimer coverings in the even
and odd sectors. The splitting between $N_e$ and $N_o$ is given by
$\Delta=Z_-/Z_+$ and is determined by the splitting of the two zero modes
(contributing to $Z_-$). This splitting is in turn determined by the
exponential tails of the zero modes away from the boundary: the exponential
dependence of the splitting on the cylinder length has the same correlation
length as the decay of the zero modes. Similarly to the case of the
torus, this exponential decay may be estimated in two ways. In the
first approach, the zero mode is constructed as a linear combination
of Green's functions localized near the boundary. 
Explicitly, we consider the zero mode $\Psi_0$ on the
``unfolded'' cylinder (Fig.~\ref{fig-split}b) and act on it
with the matrix $A_{ij}$ defined {\it on the whole plane}. The resulting
wavefunciton contains positions and amplitudes of the
sources for the Green's functions to reproduce the original
zero mode $\Psi_0$. 
This representation proves that the decay of zero modes (and hence
the splitting $\Delta$) obeys the 
estimate (\ref{torus-splitting-one}) with the minimization
performed over all possible vectors ${\bf R}$ connecting the opposite edges
of the cylinder (Fig.~\ref{fig-split}b) [see also the analogous
discussion below of the zero modes in the plane geometry]. 
Alternatively, we may represent the zero mode as a linear 
combination of plane waves quantized in the $x$-direction and decaying
in the $y$ direction, which leads to Eq.~(\ref{torus-splitting-two}).
The relation between those two estimates is the same as for the case
of the torus, with the same expression (\ref{torus-splitting-three})
valid in the regime $L_x \gg L_y \gg 1$.

We need to remark, however, that in the case of the cylinder,
the formulas (\ref{torus-splitting-one})--(\ref{torus-splitting-three})
do not properly take into account interference of tunneling trajectories
starting or ending at {\it neighboring} points of the boundary.
As a consequence, the expressions 
(\ref{torus-splitting-one})--(\ref{torus-splitting-three}) give
good (most likely exact) asymptotic estimates in the case of 
ragged boundaries, but may strongly overestimate splitting in the
case of a regular edge. For example, at the straight edge shown
in Fig.~\ref{fig-split}b, the zero mode is exactly localized
at the boundary layer of sites,
and therefore the splitting of the topological sectors on a cylinder
with such an edge is {\it exactly zero}, in contrast with
(\ref{torus-splitting-one})--(\ref{torus-splitting-three}).

Our discussion of the cylinder may be directly extended to the
plane geometry (holes in the disc), with the only difference that there
are no counterparts of Eqs.\ (\ref{torus-splitting-two}) and
(\ref{torus-splitting-three}) in this case.
Consider specifically the simplest possible geometry: a disc
with a hole in it (Fig.~\ref{fig-plane}a). 
A reference line defining the two topological sectors
is drawn to connect the two boundaries (the external boundary
and the hole boundary). Similarly to the cylindrical geometry,
the splitting of the partition functions $Z_e$ and $Z_o$ is given
by the splitting of the zero modes at the two boundaries. Our
proof of the existence of zero modes may be easily extended to show
that in the plane geometry, the hole encircling an {\it odd} number of
sites hosts a zero mode for the {\it periodic} boundary conditions
around the hole, while a hole with an {\it even} number of internal
sites has a zero mode for the {\it anti-periodic} (vison-like,
see the last section of the paper) boundary conditions. For a hole
with an {\it odd} number of internal sites, the zero mode may be constructed
as a linear combination of Green's functions with sources located
near the boundary (the proof of this statement is similar to that in the
cylindrical geometry: it follows from acting on the zero mode with the
operator $A_{ij}$ defined on the whole plane including the hole interior).
Therefore the zero mode decays with the same correlation lengths
$\xi_{\bf n}$ as the Majorana Green's function. The same result may
also be proven for a hole with an {\it even} number of internal sites:
the proof is a straightforward generalization of the theorem about the
correlation length of the vison operator proven in the last section
of the paper. Thus, for any type of hole, the splitting between
$Z_e$ and $Z_o$ is given by Eq.~(\ref{torus-splitting-one}), with the
minimization performed over all trajectories ${\bf R}$ connecting
the inner and the outer boundaries (Fig.~\ref{fig-plane}a).
 
Note however that for certain special geometries --- in particular,
those involving straight edges stretching along the $A$ directions
(along the lattice lines, 
Fig.~\ref{fig-cylinders}) --- Eq.\ (\ref{torus-splitting-one}) 
may overestimate the splitting.
An analysis of the specific case of plane geometries with only
straight edges along the $A$ directions reveals that tunneling of
visons from the straight edges is suppressed (see also a similar
remark about the zero modes at straight edges of cylinders), and
that the splitting is given by Eq.~(\ref{torus-splitting-one}) with
the tunneling trajectories $R$ drawn between {\it corners} of the boundaries
(Fig.~\ref{fig-plane}b).

\section{Local dimer correlations and ground-state energy splitting
under perturbation}

Our discussion of the splitting in the partition functions
may be extended to compute the exponentially small splitting
in correlation functions of local operators. Consider first
the simplest correlation function --- the average dimer density
on a given link $\langle n_{ij} \rangle $ on a cylinder or on a 
disc with a hole. In the $+/-$ sectors, this expectation value
may be found as the Majorana Green's function 
$G_{ij} = \langle a_i a_j \rangle$. The partition functions
and the Green's functions with periodic/antiperiodic boundary
conditions will be denoted as $Z_+$, $Z_-$, and as $G^+_{ij}$,
$G^-_{ij}$, respectively. As explained in the previous section,
with one of those boundary conditions, the matrix $A_{ij}$ has
zero modes at the two boundaries. Without loss of generality, we
assume that the zero modes appear in the ``$-$'' sector. If we consider
only one boundary (a semi-infinite cylinder or only one hole),
the zero mode at this boundary may be chosen to be real.
We denote such zero modes by
$\Psi_a(i)$ and $\Psi_b(i)$ ($a$ and $b$ label the two boundaries
near which the zero modes are localized). In a finite system,
these two wave functions are hybridized and split. The two conjugate
wave functions $\Psi_1=\Psi_a+i\Psi_b$ and $\Psi_2=\Psi_a-i\Psi_b$
have small non-zero eigenvalues $\pm i E_0$ [the value of $E_0$
is exponentially small in system size and obeys the estimates
(\ref{torus-splitting-one})--(\ref{torus-splitting-three})].
The difference between the expectation values of $n_{ij}$ in the
{\it odd} and {\it even} sectors is
\begin{align}
&\langle n_{ij} \rangle_o - \langle n_{ij} \rangle_e
= {Z_+ G_{ij}^+ - Z_- G_{ij}^- \over Z_+ - Z_- }
- {Z_+ G_{ij}^+ + Z_- G_{ij}^- \over Z_+ + Z_- } \nonumber\\
& = {2Z_+ Z_- \over Z_+^2- Z_-^2} (G_{ij}^+ - G_{ij}^-)
\approx {2Z_- \over Z_+} (G_{ij}^+ - G_{ij}^-)
\end{align}
(in the last equality we have used $Z_- \ll Z_+$).

\Figureseven

Further, the Green's functions may be rewritten as sums over
eigenvectors $G_{ij}=\sum_n \Psi_n(i) \Psi^*_n(j) / iE_n$.
The main contribution to $G_{ij}^+ - G_{ij}^-$ comes from the
zero modes $\Psi_1$ and $\Psi_2$ in $G_{ij}^-$. The exponentially
small prefactor $Z_- / Z_+$ is canceled by the exponentially
small denominator $E_0$ in $G_{ij}^-$, and to the exponential
precision we find
\begin{equation}
\langle n_{ij} \rangle_o - \langle n_{ij} \rangle_e
\sim
\Psi_a(i) \Psi_b(j) - \Psi_b(i) \Psi_a(j)
\label{correlation-splitting}
\end{equation}
Graphically, this result may be depicted as the sum of the
two diagrams in Fig.~\ref{fig-diagrams}a.

This diagrammatic approach may be extended to the even-odd
splitting in higher-order correlation functions. The diagrams
are drawn by pairing all Majorana fermions in the correlation
function (\ref{multi-point}) according to the Wick rule,
except for one pair which is connected to the boundaries.
To ``internal'' couplings, we associate the bulk Green's 
functions, while for the ``external'' lines, we associate 
the zero-mode wave functions as in (\ref{correlation-splitting})
(an example of such a diagram is shown in Fig.~\ref{fig-diagrams}b).

Furthermore, the same type of diagrammatic rules may be
derived for the case of torus, with the only difference that
the the ``external line'' must in this case be closed
and associated with the Green's function $G_{i,j+{\bf R}}$,
where the vector ${\bf R}$ winds around the torus with the
appropriate winding number (as in Fig.~\ref{fig-split}a).
An example of such a diagram is shown in Fig.\ref{fig-diagrams}c.

In the case of cylindrical and plane geometry, the smallness
of the correlation-function splitting follows from the exponential
behavior of the boundary zero modes.
Since the zero modes rapidly decay away from the boundaries,
the splitting (\ref{correlation-splitting}) is exponentially small
in the system size (at least as small as the estimates 
(\ref{torus-splitting-one})--(\ref{torus-splitting-three}) or even
smaller). The expression for the splitting may be written as
\begin{equation}
\langle n_{ij} \rangle_o - \langle n_{ij} \rangle_e
\sim
\exp\left[-{R_{\rm in}\over \xi_{\rm in}} -{R_{\rm out}\over \xi_{\rm out}}
\right],
\label{correlation-splitting-two}
\end{equation}
where $R_{\rm in}$ and $R_{\rm out}$ are the optimal tunneling
trajectories from the link $(ij)$ to the inner and outer boundaries,
and $\xi_{\rm in}$ and $\xi_{\rm out}$ are the corresponding
correlation lengths (Fig.~\ref{fig-plane}a).
If the system has a ``weak spot'' --- a preferred vison
tunneling trajectory, the correlations in the immediate vicinity
of that trajectory are most sensitive to the choice of the
topological sector. Correlations away from
the optimal trajectory have a weaker splitting between the
sectors.
On the torus, there is no ``optimal tunneling trajectory'' because
of the translational symmetry, and the correlation-function
splitting is given by 
(\ref{torus-splitting-one})--(\ref{torus-splitting-three}),
independently of the position of the involved operator.

The above results allow us to estimate the energy splitting in the
perturbed Rokhsar--Kivelson model to the first order of the
perturbation theory. Physically, two types of perturbations may
be of interest. First, we consider the case $v/t<1$ in the
Hamiltonian (\ref{dimer-hamiltonian}), in which case the ground-state
energy is no longer zero. The ground-state energy may be estimated
to the linear order in $(1-v/t)$ as
$(v-t)\langle H_v \rangle_{RK}$, where $H_v$
is the potential (proportional to $v$) term in the 
Hamiltonian (\ref{dimer-hamiltonian}), and the average is taken
in the unperturbed system. $H_v$ is a four-point (dimer-dimer)
correlation function, and from our previous discussion it
follows that the splitting of $\langle H_v \rangle_{RK}$ in
different topological sectors is exponentially small in the
system size with the exponent estimated by
(\ref{torus-splitting-one})--(\ref{torus-splitting-three}).
Second, similar exponential smallness may also be derived for the 
energy splitting under the influence of disorder (again, to the
linear order in the disorder potential). Note that the energy splitting
is more sensitive to disorder in the vicinity of the optimal vison
tunneling trajectory.

Within our method, the exponentially small response 
of the energy splitting to the perturbation can be derived only to the
lowest order of the perturbation theory. Extending those results to
higher orders requires
information on the excited states at the RK point. This goes beyond the
scope of this paper, since the excited states are not accessible by the
Pfaffian technique.

\section{Vison operators and correlations}
\label{section-vison}

The concept of a vison can be made more transparent by introducing a
``vison operator''. Given two elementary plaquets (triangles) of the
lattice and a contour $\Gamma$ connecting them, we can define the
product of two vison operators $V_1 V_2$ as
\begin{equation}
V_1 V_2 = (-1)^{N_\Gamma} = \prod_{\Gamma} (1-2 n_{ij}),
\label{vison-vison}
\end{equation}
where $N_\Gamma$ is the number of dimers intersecting the contour $\Gamma$,
and the last product is taken over all links $(ij)$ intersecting $\Gamma$
(Fig.~\ref{fig-visons})~\cite{Read-Chakraborty}.
It can be easily seen that deforming the contour $\Gamma$ while keeping
the end points fixed may only change the sign of this operator (depending
on whether between the old and the new contour lies odd or even number of 
sites). This allows us to define a single vison operator (up to a sign;
more precisely, the vison operator is defined on the frustrated dual
lattice, with reversing its sign when moved around one site of the
original lattice). For a single vison operator, the contour $\Gamma$
is drawn either to infinity (for an infinite system) or to the
boundary. In a finite system without boundaries (e.g.\ torus), vison
operators may appear only in pairs.

\Figureeight

In the Majorana fermion technique (\ref{pfaffian}), the vison operator
is equivalent to changing the boundary conditions (from periodic to
anti-periodic and vice versa, i.e., placing a $Z_2$ twist) across the
contour $\Gamma$. Thus switching topological sectors on a cylinder as
described above is equivalent to placing a pair of visons in the
cylinder openings. On a torus, switching topological sectors is achieved
by creating a vison-vison pair, subsequently moving one of the two
visons around the torus and annihilating it back with the other vison.

Applying a vison operator as defined above to the ground state does not
produce an eigenstate of the Hamiltonian (\ref{dimer-hamiltonian}). However,
preliminary studies show that the state thus obtained is close to the
actual low-lying excitations\cite{in-preparation}. A localized
excitation (a wavepacket)
may be classified as either vison-like or non-vison-like
depending on whether it originates a cut $\Gamma$ across which the boundary
conditions are changed. Based on the available numerical 
studies \cite{in-preparation}, we believe that the lowest excitations
belong to the vison-like sector; this also conforms to the existing
field-theoretical description of the RVB liquid \cite{Senthil-Fisher}
(studies of related models with an exact construction of the vison 
excitations appeared recently in 
Refs.~\onlinecite{Misguich-Serban-Pasquier,Ioffe-Feigelman}).
We expect that the leading exponential asymptotics of long-distance
correlators of vison-like operators is dominated by the topological
effect of changing the boundary conditions across the cut $\Gamma$, and
only finite prefactors depend on the details of the involved operators.
In other words, the correlation length of vison-like operators
is universal and may be determined, for example, from the pairwise
correlation function $\langle V_1 V_2 \rangle$ as defined in 
(\ref{vison-vison}). 

We can prove that the vison correlation length equals that for
Majorana fermions. For the same reason as for a hole in the
plane (see discussion above), a vison on an infinite plane
produces a zero mode of the matrix $A_{ij}$. Thus the vison
correlation length is determined by the decay of this zero mode
at large distances. The matrix $A_{ij}$ in the presence of the
vison may be written as $A_{ij}=A_{ij}^{(0)}+\delta A_{ij}$,
where $A_{ij}^{(0)}$ is the matrix of hopping amplitudes without the vison
and $\delta A_{ij}$ is localized at the cut drawn from the vison.
Note that the zero mode does not depend on the trajectory of the cut
(up to a gauge transformation). The equation on the zero mode
$(A^{(0)}+\delta A) \Psi_0 =0$ may be rewritten as
$\Psi_0= -[A^{(0)}]^{-1} \delta A \Psi_0$. From this expression,
it follows that the exponential decay of $\Psi_0$ away from the
vison has the asymptotics of $ [A^{(0)}]^{-1}$ which is
exactly the Green's functions of Majorana fermions 
(\ref{Majorana-Green}). This finishes the proof.
Note that this result is consistent with our findings
for the splitting between topological sectors and with their
interpretation in terms of vison tunneling amplitudes. Furthermore,
our result about the decay of the zero mode at the point-like vison 
is a particular 
case of a more general statement about the zero modes at even-size holes
(see our discussion in Section~\ref{section-degeneracy}): the point-like
vison may be considered as a hole of size zero; the proof may be
extended in a straightforward way to cover the general case of
an even-size hole.

\Figurenine

To illustrate our result on the equality between the correlation
lengths for visons and
for Majorana fermions, we have calculated numerically the
vison-vison correlation function in the $A$ direction. The calculation
was performed by the variational Monte Carlo method: the vison
correlation function was computed as the average over a suitable random
walk in the space of all dimer configurations\cite{Balents-Fisher-Girvin}. 
Figure~\ref{fig-numerics} shows
the numerical results of the calculation (for distances up to 9 lattice
spacings, on the lattice of 50$\times$50 torus), together with the
analytic expression $R^{-1/2}\exp(-\Re\, R/\xi_A)$. We have no analytic
expression for the phase of the oscillating part of the correlation function 
(arising from the imaginary part of $\xi_A^{-1}$) and
therefore compare only the overall magnitude of the correlation function. 
The agreement in the magnitude and the change of sign in the vison correlator
(a signature of the oscillatory behavior) are consistent with our
analytical prediction of the correlation length (\ref{xi-A}).

\bigskip

To summarize, we have shown that in the
dimer model (\ref{dimer-hamiltonian}) on the triangular lattice,
at the RK point $t=v$, both
dimer and vison correlations decay exponentially with the distance.
Dimer correlations may be expressed as finite products of the Green's
functions of auxiliary Majorana fermions for which the correlation length
is computed explicitly. We have further shown that the vison correlation length
coincides with that of Majorana fermions. The same correlation length
also governs the splitting between correlation functions in different
topological sectors
(\ref{torus-splitting-one})--(\ref{torus-splitting-three}).
Thus the Rokhsar--Kivelson dimer model provides an example of the
system where the topological ground-state properties of the RVB liquid may
be explicitly verified.

We thank L.~Ioffe, G.~Blatter,
T.~Senthil, and O.~Motrunich for useful discussions. 
Support from the program ``Quantum Macrophysics'' of 
Russian Academy of Sciences, the program
``Physics of Quantum Computations'' of Russian Ministry of Science,
SCOPES program of Switzerland, Dutch Organization for Fundamental 
research (NWO), RFBR grant 01-02-17759, 
and the Swiss National Foundation is gratefully acknowledged.

{\bf P.\ S.\ } At the final stage of the manuscript preparation, we
have learned about the recent work \cite{Fendley}, where the same
dimer model is studied. Many of our results in Section~\ref{section-pfaffian}
overlap with those derived in that work.
Among other results, Ref.~\onlinecite{Fendley} also contains a 
more detailed analysis of the Majorana Green's functions.


\begin{thebibliography}{}

\bibitem{PWA} P.~W.~Anderson, Science {\bf 235}, 1196 (1987).

\bibitem{Kivelson-Rokhsar-Sethna}
S.~A.~Kivelson, D.~S.~Rokhsar, and J.~P.~Sethna,
Phys.\ Rev.\ B {\bf 35}, 8865 (1987).

\bibitem{Fradkin-Kivelson}
E.~Fradkin and S.~Kivelson,
Mod.~Phys.~Lett.\ B {\bf 4}, 225 (1990).

\bibitem{Senthil-Fisher}
T.~Senthil and M.~P.~A.~Fisher,
Phys.\ Rev.\ B {\bf 62}, 7850 (2000);
ibid. {\bf 63}, 134521 (2001).

\bibitem{Lhuillier}
G.~Misguich et al, Phys.~Rev.\ B {\bf 60}, 1064 (1999);
R.~Sindzingre, C.~Lhuillier, and J.-B.\ Fouet, cond-mat/0110283;
G.~Misguich et al, cond-mat/0112360.

\bibitem{Moessner-Sondhi}
R.~Moessner and S.~L.~Sondhi, Phys.\ Rev.\ Lett.\ {\bf 86}, 1881 (2001).

\bibitem{Balents-Fisher-Girvin}
L.~Balents, M.~P.~A.~Fisher, and S.~M.~Girvin,
cond-mat/0110005.

\bibitem{Senthil-Motrunich}
T.~Senthil and O.~Motrunich, cond-mat/0201320;
O.~Motrunich and T.~Senthil, cond-mat/0205170.

\bibitem{Misguich-Serban-Pasquier}
G.~Misguich, D.~Serban, V.~Pasquier, cond-mat/0204428.

\bibitem{Ioffe-Feigelman}
L.~B.~Ioffe and M.~V.~Feigelman, cond-mat/0205186.

\bibitem{Moessner-Sondhi-Fradkin}
R.~Moessner, S.~L.~Sondhi, and E.~Fradkin,
cond-mat/0103396.

\bibitem{Read-Sachdev}
N.~Read and S.~Sachdev, Nucl.~Phys.\ B {\bf 316}, 609 (1989). 

\bibitem{LDA}
S.~Liang, B.~Doucot, and P.~W.~Anderson,
Phys.\ Rev.\ Lett.\ {\bf 61}, 365 (1988);
B.~Sutherland, Phys.~Rev. B {\bf 37}, 3786 (1988).

\bibitem{Read-Chakraborty}
N.~Read and B.~Chakraborty, Phys.\ Rev.\ B {\bf 40}, 7133 (1989).

\bibitem{Rokhsar-Kivelson}
D.~S.~Rokhsar and S.~A.~Kivelson,
Phys.\ Rev.\ Lett.\ {\bf 61}, 2376 (1988).

\bibitem{Bonesteel}
N.~Bonesteel, Phys.\ Rev.\ B {\bf 40}, 8954 (1989).

\bibitem{Levitov}
L.~S.~Levitov, Phys.\ Rev.\ Lett.\ {\bf 64}, 92 (1990).

\bibitem{Ioffe}
L.~B.~Ioffe et al, Nature {\bf 415}, 503 (2002).

\bibitem{Kasteleyn}
P.~W.~Kasteleyn, J.~Math.~Phys.\ {\bf 4}, 287 (1963);
S.~Samuel, J.~Math.~Phys.\ {\bf 21}, 2806 (1980).

\bibitem{in-preparation}
L.~B.~Ioffe, unpublished.

\bibitem{Fisher-Stephenson}
M.~E.~Fisher and J.~Stephenson,
Phys.~Rev.\ {\bf 132}, 1411 (1963).

\bibitem{I-S}
D.~A.~Ivanov and T.~Senthil, cond-mat/0204043.

\bibitem{Zecchina}
T.~Regge and R.~Zecchina, J.~Phys. A {\bf 33}, 741 (2000)
[cond-mat/9909168];
R.~Zecchina, Physica A, {\bf 302}, 100 (2001).

\bibitem{Read-Green}
N.~Read and D.~Green,
Phys.\ Rev.\ B {\bf 61}, 10267 (2000)

\bibitem{Iv}
D.~A.~Ivanov, Phys.~Rev.~Lett.\ {\bf 86}, 268 (2001).

\bibitem{Fendley}
P.~Fendley, R.~Moessner, and S.~L.~Sondhi, cond-mat/0206159.

\end{thebibliography}
\end{document}